\documentclass[prb,preprint]{revtex4}

\pdfoutput=1

\usepackage{graphicx}

\begin{document}

\title{Virial ratios of Kepler motion}

\medskip 

\date{October 1, 2024} \bigskip

\author{Manfred Bucher \\}
\affiliation{\text{\textnormal{Physics Department, California State University,}} \textnormal{Fresno,} \textnormal{Fresno, California 93740-8031} \\}

\begin{abstract}
Recently it was shown that the ratio of kinetic energy $K$ and potential energy $U$ at the perihelion of a Kepler orbit relates to the ellipse's eccentricity, $-2K/U = 1 + e$. Here, a general expression for the virial ratio at {\it any} position on the orbit is presented and its relation to the virial theorem is revealed.
\end{abstract}

\maketitle

It is amazing that 400 years after Kepler's discovery of his eponymous laws, new aspects of planetary motion are still found.\cite{1,2,3,4,5,6} Recently it was shown that the ratio of kinetic energy $K(\mathbf{r})$ and potential energy $U(\mathbf{r})$ at the perihelion $\mathbf{A}$ of a Kepler orbit,
\begin{equation}
    \rho(\mathbf{r_A}) = \frac{2K(\mathbf{r_A})}{-U(\mathbf{r_A})} = 1 + e \;,
\end{equation}
relates to the ellipse's eccentricity $e$.\cite{7} Using data of the mass $m$, closest distance $r_\mathbf{A}$, and closest approach velocity $v_\mathbf{A}$, the author calculates with Eq. (1) the eccentricity of planets and comets to high agreement with published data.\cite{7} The author also points out the similarity of Eq. (1) with the virial theorem, here for motion subject to an inverse-square central force, $F \propto r^{-2}$,
\begin{equation}
     <\rho> \; = \frac{2<K>}{-<U>} = 1 \;,
\end{equation}
which relates the {\it average} kinetic and potential energy of an orbiting body. In this note, a general expression for the viral ratio at any position on the orbit is presented and its relation to the virial theorem is shown.

With the assumption that the central mass $M$ is much larger than the orbiting mass $m$, $M>>m$, the (conserved) total energy $E$ of an orbiting body at any position $\mathbf{r}$ of the orbiting body is
\begin{equation}
 E = -\frac{mMG}{2a} = K(\mathbf{r}) + U(\mathbf{r}) = K(\mathbf{r}) -\frac{mMG}{r} \;,
\end{equation}
where $G$ is the universal gravitational constant and $a$ the semimajor axis of the orbital ellipse. Rearrangement of terms gives the {\it local} virial ratio, 
\begin{equation}
    \rho(\mathbf{r}) = \frac{2K(\mathbf{r})}{-U(\mathbf{r})} = 2 - \frac{r}{a} \;.
\end{equation}
For the perihelion $\mathbf{A}$, $r = a - c$, with $c = \sqrt{a^2 - b^2}$ being the distance from the ellipse's focus $\mathbf{F}$ to center $\mathbf{C}$ --- see Fig. 1 --- and eccentricity $e = c/a$, Eq. (1) is recovered.
For the aphelion $\mathbf{A'}$, where $r = a + c$, the local virial ratio is
\begin{equation}
    \rho(\mathbf{r_{A'}}) = 1 - e \;,
\end{equation}
in opposite deviation from unity as at the perihelion $\mathbf{A}$.
At the intersection of the minor axis with the orbit, $\mathbf{B}$, where $r = a$,
 \begin{equation}
    \rho(\mathbf{r_B}) = 1 \;,
\end{equation}

\pagebreak
\footnotesize 
\noindent FIG. 1. Ellipse geometry
\normalsize

\includegraphics[width=6in]{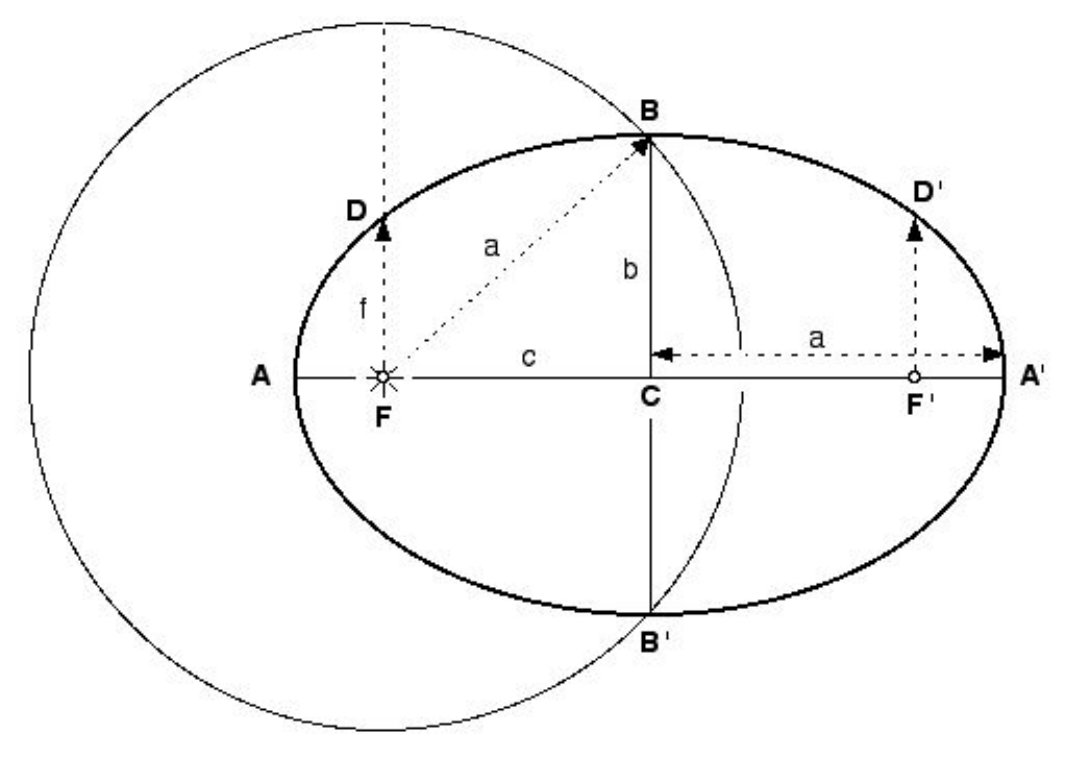} 

\noindent as for the average relation of the viral theorem, Eq. (2). 
Another prominent position of the orbit is at the intersection of the focal axis with the orbit --- position $\mathbf{D}$. With distance $r = f = b^2/a$, called ``semi latus rectum,'' Eq. (4) gives
\begin{equation}
    \rho(\mathbf{r_D}) = 1 + e^2 \;,
\end{equation}
a smaller value than at the perihelion, Eq. (1). For the position $\mathbf{D'}$ --- symmetric to $\mathbf{D}$ with respect to the minor axis, but related to the \emph{empty} focus $\mathbf{F'}$ --- at distance $r = 2a - f$ from focus $\mathbf{F}$, we obtain
\begin{equation}
    \rho(\mathbf{r_{D'}}) = 1 - e^2 \;,
\end{equation}
being of larger value than at the aphelion, Eq. (5).

From Eq. (4) one intuits an odd symmetry of the local virial ratio $\rho(\mathbf{r})$ with respect to the minor axis $BB'$ and with extremes at the perihelion $\mathbf{A}$ and aphelion $\mathbf{A'}$, causing mutual cancellations that lead to the average ratio of the virial theorem, Eq. (2). To show the cancellation, we integrate Eq. (4) over a symmetric half of the orbit --- say, from the perihelion $\mathbf{A}$ to the lateral point $\mathbf{B}$, 
\begin{equation}
<\rho>_{A}^B = \frac{\int_{a-c}^{a} 2 - \frac{r}{a} \,dr }{\int_{a-c}^{a}  \,dr } = 1 +\frac{a-f}{2c}  \:,
\end{equation}
and from point $\mathbf{B}$ to the aphelion,
\begin{equation}
<\rho>_B^{A'} = \frac{\int_{a}^{a+c} 2 - \frac{r}{a} \,dr }{\int_{a}^{a+c}  \,dr } = 1 - \frac{a-f}{2c} \:,
\end{equation} 
with equal and opposite contributions that indicate the dominance of kinetic over potential energy or vice versa.

\end{document}